# Altermagnetism with Non-collinear Spins

Sang-Wook Cheong and Fei-Ting Huang

Rutgers Center for Emergent Materials and Department of Physics and Astronomy, Rutgers University

*Corresponding author: sangc@physics.rutgers.edu

**ABSTRACT**

**Altermagnetism is introduced as a category of magnetic states with vanishing net magnetic moment, and consists of 'collinear' alternating (i.e., antiferromagnetic-like) spins and alternating variations of local structures around spins in such a way that the symmetry allows ferromagnetic behaviors such as anomalous Hall effect and magneto-optical Kerr effects. Altermagnets exhibiting ferromagnetic behaviors without any external perturbations (called type-I) turn out to belong to the ferromagnetic point group, and represent a form of weak ferromagnetic states. Other altermagnets (called type-II and type-III) can have ferromagnetic behaviors only with external perturbations, which conserve parity-time reversal (PT) symmetry. All types of altermagnets themselves have broken PT symmetry. The concept of altermagnetism can be extended to accommodate multiple spin and local-structure variations (thus, include 'non-collinear' spins), and this extended form of altermagnetism offers an intriguing opportunity to leverage complementary advantages of both ferromagnetism and antiferromagnetism, and thus holds great promise for, for example, various spintronic applications.**

# INTRODUCTION

Spintronics[1-4] where spin and charge degrees of freedom are mutually coupled and manipulated through crossing conjugate fields (i.e. electric fields for spins and magnetic fields for charges) has been an active research area for the last three decades since the discovery of giant magnetoresistance[5-8], and has been well implemented in real devices. Traditionally, spintronics utilizes ferro-(ferri)magnets since the manipulation and detection of these ferro-(ferri)magnetic states are straightforward. But antiferromagnetic spintronics[1, 3, 9-11] has become highly topical due to the active manipulation of the antiferromagnetic state and its magnetic textures via spin and charge currents. Antiferromagnetic materials, in general, could embody the numerous interesting features beneficial for spintronic applications[12-14]: they produce no stray fields, so are robust against external magnetic fields and make them suitable for device miniaturization, display ultrafast dynamics in THz ranges, and are capable of generating good spin-current transport with micrometer spin-diffusion lengths.

Altermagnets[15-20] are magnets with collinear antiferro-arrangement of one-kind spins (i.e. 'alter'nating spins), and also simultaneously with 'alter'nating orientations of local structures around spins in such a way that the symmetry allows ferromagnetic behaviors (non-zero net magnetic moment, Anomalous Hall effect (AHE), etc.)[18, 21]. In other words, in altermagnets, spin magnetic moments are fully compensated when spin-orbital coupling (SOC) is zero, but non-zero SOC, enabling coupling between spins and alternating local structures (such as oxygen coordination around magnetic ions), can result in a non-zero net magnetic moment and other ferromagnetic behaviors through uncompensated 'orbital magnetic moments'.

# FERROMAGNETISM V.S. ANTIFERROMAGNETISM

Surprisingly, antiferromagnetism is not formally defined even though the term has been used frequently, but the ferromagnetic point group is well defined in terms of symmetry. Magnetization (*M*) along *z* has broken {$I\otimes T, T, M_x, M_y, R_x, R_y, C_{3x}, C_{3y}$} with free rotation along *z* (see Ref. 22-23 for the definition of symmetry operation notations[22,23]). All magnetic point groups (MPGs), belonging to the ferromagnetic point group[22,24], do have, at least, broken {$I\otimes T, T, M_x, M_y, R_x, R_y, C_{3x}, C_{3y}$} with free rotation along *z* or the relevant requirements along *x* or *y*, i.e., have symmetry operational similarity (SOS) with *M* [23,25,26]. The thirty one (31) ferromagnetic MPGs[24,27] include 1, $\bar{1}$, 2, 2′, *m*, *m*′, 2/*m*, 2′/*m*′, 2′2′2, *m*′*m*′*m*, *m*′*m*′2, *m*′*m*2′, 4, $\bar{4}$, 4/*m*, 42′2′, 4*m*′*m*′, $\bar{4}$2′*m*′, 4/*mm*′*m*′, 3, $\bar{3}$, 32′, 3*m*′, $\bar{3}$*m*′, 6, $\bar{6}$, 6/*m*, 62′2′, 6*m*′*m*′, $\bar{6}$*m*′2′, 6/*mm*′*m*′. We have considered symmetries that are along only those basis vectors of the conventional crystallographic coordinate systems. Those basis vectors are given in the settings of 122 MPGs. For example, we discuss the symmetries along *x*, *y*, *z*, *xy*, and *yx* directions of the tetragonal and cubic MPGs, while only symmetries along *x*, *y*, *z* directions in orthorhombic MPGs are considered in our SOS analysis. Even for hexagonal and trigonal structures, *x*, *y*, and *z* are defined to be orthogonal to each other. The rotation $C_3$ encompasses both $C_3^+$ (counterclockwise) and $C_3^-$ (clockwise), with the + and − signs. Any symmetry criteria associated with $C_3$ are equally applicable to the operation on $C_3^+$ and $C_3^-$, concurrently. This principle extends to $C_4$ and $C_6$ rotations.

It is rigorous and produces no ambiguity to define ferromagnetism and antiferromagnetism in crystalline materials in this manner: any magnetic states, whose MPGs belong to the ferromagnetic point group, are ferromagnetic; otherwise, antiferromagnetic[24,28,29]. These ferromagnetic MPGs do exhibit unique physical properties such as non-zero net magnetic moment, nonreciprocal Faraday optical rotation, and linear anomalous Hall effect (AHE). (Note that there are four types of anomalous Hall effects (true anomalous Hall, Ettingshausen, Nernst, and thermal Hall effects), and in terms of symmetry, there is little difference among the requirements for four types of anomalous Hall effects as long as the relevant electric/thermal current can exist[22].) The requirement to observe each of the above phenomena is identical with that for belonging to the ferromagnetic point group, so these effects can be observed 'only' in ferromagnetic MPGs. Ferromagnets also exhibit diagonal piezomagnetic effects along the net magnetic moment directions and magneto-optical Kerr effects (MOKEs); however, the diagonal piezomagnetic effect and MOKE can be also observed in some antiferromagnets with properly-broken symmetries[22, 30, 31]. Simple ferromagnets with all ionic magnetic moments pointing one direction do belong to ferromagnetic point groups, but ferrimagnetic, canted antiferromagnetic, weak ferrimagnetism or weak ferromagnetic states with non-zero net magnetic moments all also belong to the ferromagnetic point group. For the sake of simplicity, we will classify all magnets, belonging to the ferromagnetic point group, but not exhibiting simple ferromagnetism or ferrimagnetism (having uncompensated moments with antiferro-arranged multiple-kinds of ionic magnetic moments), as 'weak ferromagnets'[24, 28]. We highlight that antiferromagnets are magnetic states found in crystalline solids, which do not belong to the ferromagnetic point group.

**TYPE-I AND TYPE-II ALTERMAGNETISM**

PT symmetry (P; parity, T; time reversal, and PT; parity times time reversal) is broken in all altermagnets. The essential aspect of altermagnets is: when spin-orbit coupling is ignored, the spin angular momenta in an altermagnet are fully compensated, so there is no net spin angular momentum, but the spin lattice with crystallographic variations have broken PT symmetry, so there exists 'spin split in excitations' and 'non-zero net magnetic momentum through spin-orbit coupling in zero external perturbations or under PT-symmetric external perturbations'. Note that the crystallographic lattices of altermagnets are often centrosymmetric (i.e., PT symmetric), but can be also non-centrosymmetric (i.e., PT-symmetry broken). The prominent effect of PT symmetry breaking is to create nontrivial magnetism-related physics by lifting the Kramers degeneracy and producing spin textured electronic bands, i.e., band-dependent spin directions. Note that PT symmetry can be broken in three distinct manners: [1] broken P, but unbroken T, [2] broken T, but unbroken P, and [3] broken P, broken T, and PT is also broken if P and T are broken in different ways. For example, the necessary condition for odd-order AHE, which includes linear ($1^{st}$-order) AHE, is broken PT symmetry[22]. Precisely speaking, there can be two different types of altermagnets: type-I has non-zero net magnetic moment and ferromagnetic behaviors without external perturbations, and type-II has zero net magnetic moment and no ferromagnetic behaviors without the external perturbation but exhibit non-zero net magnetic moment and ferromagnetic behaviors in the presence of external PT-symmetric perturbations such as applied electric/thermal current, light illumination or stress. It turns out that both type-I and type-II altermagnets do have the broken PT symmetry. Note that external perturbations such as applied electric/thermal currents, light illumination or stress themselves do not break the PT symmetry, but, for example, applied electric fields or magnetic fields do break the PT symmetry. We also emphasize that the entire experimental set-up, combining a type-II altermagnet with the broken PT symmetry and an external perturbation with the unbroken PT symmetry, now has SOS with *M*, so it can exhibit ferromagnetic behaviors. Figs. 2a-c exemplify type-I altermagnets, and we call alternating 2-fold-symemtric local structures around spins as directors (shown as gray ellipsoids in Fig. 2). All type-I altermagnets turn out to be weak ferromagnets. All MPGs for altermagnetism with collinear spins[18] identified in Ref 18 include $1, \bar{1}, 2, 2', m, m', 2/m, 2'/m', 2'2'2, m'm'm, m'm2'$, and $m'm'2$ (inside of the dashed circle in Fig. 1). The altermagnets of Fig. 2a-c have $m'm'm$-type MPGs.

Magnetic anisotropy often plays an important role in understanding altermagnetism and weak ferromagnetism, in general. There exist three different manners to have magnetic anisotropy: single ion anisotropy, anisotropic symmetric or antisymmetric exchange coupling and g-tensor anisotropy (i.e., orientation-dependent magnitude of the total magnetic moment)[32, 33]. The origin of net magnetic moment in altermagnets in Fig. 2a and b is g-tensor anisotropy – in other words, the true spins in those states are fully antiferromagnetic (i.e., fully compensated), but their symmetry allows non-zero net orbital magnetic moments through SOC. The net canted moment in Fig. 2c, where pure spin moments are 45° degree away from the *x* and *y* axes, is due to the Dzyaloshinskii–Moriya (DM) interaction (i.e., antisymmetric exchange coupling), rather than g-tensor anisotropy, from SOC. Note that both Dzyaloshinskii–Moriya interaction and g-tensor anisotropy originate from SOC. Fig. 2b corresponds to, for example, one possible magnetic state (not ground state) in $RuO_2$[18] and one possible magnetic state in MnTe[21, 34], and Fig. 2c represents, for example, the magnetic state[27] of $Gd_2CuO_4$, the magnetic state of each layer of $La_2CuO_4$ [5] or $Sr_2IrO_4$,[35] and the magnetic state of NbMnP [36].

Type-II altermagnets do not have non-zero net magnetic moment nor exhibit ferromagnet-like behaviors but still have the broken PT symmetry. One example of type-II altermagnets is shown in Fig. 3c with zero net moment but the broken PT symmetry, which corresponds to $4'/mm'm$. It turns out that these type-II altermagnets can still exhibit ferromagnet-like behavior in the presence of applied current or uniaxial stress (conserving the PT symmetry), so, for example, off-diagonal net moment can be induced with even-order of applied electric current, AHE can be induced in high-odd-order ($3^{rd}$-, $5^{th}$-order, etc.) with applied electric current, and off-diagonal net moment can be induced by uniaxial stress. MPGs for high-odd-order AHE, off-diagonal even-order current-induced magnetization, and off-diagonal piezomagnetism, include 32, 3$m$, $\bar{3}m$, 4′, $\bar{4}$′, 4′/$m$, 4′22′, 4′$m'm$, $\bar{4}$′2′$m$, $\bar{4}$′2$m$′, 4′/$mmm$′, 6′, $\bar{6}$′, 6′/$m$′, 6′22′, 6′$mm$′, $\bar{6}$′$m$2′, $\bar{6}$′$m$′2, 6′/$m'mm$′, 23, $m\bar{3}$, 4′32′, $\bar{4}$′3$m$′, and $m\bar{3}m$′ (see Ref. 22). Note that the $m'mm$′ state in Fig. 2b & c can exhibit net magnetic moment along $y$, linear AHE with current along $x$ (or $z$) and Hall voltage along $z$ (or $x$), off-diagonal piezomagnetism with uniaxial stress along $x$ or $z$ and induced magnetization along $y$, and also diagonal piezomagnetism along $y$. The 4′/$mm'm$ state in Fig. 3c can exhibit high-odd-order AHE with current along $x$ (or $y$) and Hall voltage along $y$ (or $x$), even-order current-induced magnetization with current along $x$ or $y$ and induced net moment along $z$, and off-diagonal piezomagnetism with uniaxial stress along $x$ or $y$ and induced net moment along $z$. Note that Fig. 3d is also a type-II altermagnet with 4′/$mmm$′, and corresponds to CoF$_2$, which is well known to exhibit off-diagonal piezomagnetism[37, 38].

**ALTERMAGNETISM WITH NON-COLLINEAR SPINS**

Altermagnetism does not have to have only 2 alternating directors and collinear antiferromagnetic spins (i.e., 2 spin orientations). One can expand the concept of altermagnetism to include multiple directors (in general, local structural parameters such as directors, pseudo-scalars, electric polarizations, etc.) and non-collinear spins. Examples of altermagnetism with the extended definition and one-kind magnetic ions are listed in Figs. 2d-g and Fig. 3a-b. We emphasize that in all cases in Fig. 2 and Fig. 3, pure spin moments without SOC are fully compensated. An altermagnetic state with 3 different director orientations and 3 different spin orientations on a triangle is displayed in Fig. 2d, and the spin configuration in Kagome lattice with 120°-ordered spins, shown in Fig. 2e, can be considers as a combination of the Fig. 2d state and the mirror image of the Fig. 2d state (the relevant mirror is perpendicular to the $x$ axis). Note that the butterfly tie pattern (two gray triangles) around magnetic ions in Fig. 2e acts as a director, and this Fig. 2e state represents, for example, the magnetic state in Mn$_3$Ge (*mm'm'*)[13, 39, 40]. Fig. 2f depicts a monopole-type 120° spin order with an alternating arrangement of 3 different director orientations on a trimerized triangular lattice, which corresponds to the MPG of $\bar{6}'2m'$. Figs. 2g & 2h show altermagnetic states with 3 different director orientations and 3 different spin orientations in a trimerized triangular lattice. The Fig. 2g state with 3*m'* (polar and ferromagnetic) is relevant to, for example, the A$_2$ phase in hexagonal REMnO$_3$ (RE=rare earths)[41, 42] or (Lu,Sc)FeO$_3$ [43]. The Fig. 2h state with 32' (chiral and ferromagnetic) is relevant to, for example, a proposed magnetic state of chiral Co$_{1/3}$TaS$_2$ [44]. We emphasize that the crystallographic lattices of Figs. 2a-c, & e and Figs. 3a-g are centrosymmetric (i.e. PT symmetric), but those of Figs. 2f-h and Figs. 3h-i are non-centrosymmetric (i.e., PT-symmetry broken). Figure 3a depicts an altermagnetic state with 4 different director orientations and 4 different spin orientations in a square lattice. All these extended altermagnets with fully compensated spins are weak ferromagnets, and g-tensor anisotropy and DM interaction from SOC is responsible for net magnetic moments in Figs. 2a, b, d, e, & Fig. 3a, and Figs. 2c, g, & h, respectively. Note that when g-tensor anisotropy, which is usually very small, produces a tiny net magnetic moment in, for example, Figs. 2a, b, d, e, & Fig. 3a. However, other ferromagnet-like physical effects such as linear AHE and Faraday rotation can be significant through mechanisms associated with the Berry curvature. In addition, since the most part of magnetic moments in the altermagnets are antiferro-type, the magnetic dynamics of altermagnets can be fast like that in typical antiferromagnets. When net magnetic moment is reversed in all (extended) altermagnets in Figs. 2a-h by, for example, applied magnetic fields, the entire spins rotate by 180°, i.e. there is a switching between time-reversal domains, even though the net magnetic moment can be very small and most parts of magnetic moments are antiferro-arranged. Thus, various ferromagnetic-like behaviors such as linear AHE and MOKE can also be switched with the flipping of the tiny net magnetic moments.

Finally, we note these important aspects: [1] it is not necessary to have the same number of director orientations and spin orientations for the extended altermagnetism, even though the number of director orientations (or pseudo-scalars) and that of spin orientations in all (extended) altermagnet examples in Fig. 2 and Figs. 3a, c & d are identical. One example is an altermagnet with 2 spin orientations with 3 director orientations in a square lattice (see Fig. 3b), whose net magnetic moment is due to g-tensor anisotropy. [2] Type-II altermagnetism can be also achieved with the combination of 3 directors and 120° spin order as shown in Fig. 3e with stacked trimerized triangular lattice. This type-II altermagnet with the broken PT symmetry corresponds to the MPG of $6'/m'm'm$, and is not associated with any net magnetic moment, but can exhibit, for example, high-odd-order AHE and off-diagonal piezomagnetism. On the other hand, the magnetic states in Figs. 3f & 3g with stacked trimerized triangular lattice, 3 directors and 120° spin order have the unbroken PT symmetry, so cannot exhibit any kind of (current/stress-induced) ferromagnetic behaviors; for example, any AHE is absent. [3] Instead of being limited to mere alternation of director orientations or pseudo-scalars, the extended altermagnetism can encompass various types of alternations in the relevant local structures surrounding spins. These may include changes in the magnitudes of directors, anion coordination or the orientation of local polar structures. For example, Figs. 3h & 3i depict collinear magnetic state on stacked honeycomb lattice with two types of oxygen coordination (small gray triangles: tetrahedral coordination; large gray circles: octahedral coordination). The magnetic state in Fig. 3h with $6m'm'$ is a type-I altermagnet that can exhibit linear AHE, and represents the magnetic state of polar $Mn_2Mo_3O_8$ [45]. In contrast, the magnetic state in Fig. 3i with $6'mm'$ turns out to be a type-II altermagnet, and represents the magnetic state of polar $Fe_2Mo_3O_8$ [46]. The magnetic state in Fig. 3i can show high-odd-order AHE with current along *y* and Hall voltage along *z*, even-order current-induced magnetization with current along *y* and induced magnetization along *x*, Off-diagonal piezomagnetism with stress along *y* and induced magnetization along *x*.

Here is the summary list of all altermagnets that we have discussed: (1) Type-I altermagnets: MPG *m'mm'* as shown in Figs. 2a-c, Fig. 2e, and Fig. 3b; MPG 2*m'm'* in Fig. 2d; MPG 3*m'* in Fig. 2g; MPG 32' in Fig. 2h; MPG 2'/*m'* in Fig. 3a; MPG 6*m'm'* in Fig. 3h. (2) Type-II altermagnets: MPG $\bar{6}'2m'$ in Fig. 2f; MPG 4'/*mmm'* in Figs. 3c-d, MPG 6'/*m'm'm* in Fig. 3e, and MPG 6'*mm'* in Fig. 3i.

We have discussed altermagnets that can exhibit odd-order AHE - either linear AHE with non-zero (zeroth-order) net magnetic moments in Type-I altermagnets or high-odd-order AHE with off-diagonal even-order current-induced magnetic moments in Type-II altermagnets. However, broken PT symmetry encompasses more than just odd-order AHE. There can be altermagnets with broken PT symmetry, which do not exhibit odd-order AHE – these can be coined Type-III altermagnets. The relevant MPGs include 11', *m*1', 21', *mmm*, *mm*2, *mm*21', 222, 2221', 31', 3*m*1', 321', 41', $\bar{4}$1', 422, 4221', 4*mm*, 4*mm*1', $\bar{4}$2*m*, $\bar{4}$2*m*1', 4/*mmm*, 61', 6*mm*, 6*mm*1', $\bar{6}$1', $\bar{6}m$2, $\bar{6}m$21', 622, 6221', 6/*mmm*, 231', 432, 4321', $\bar{4}$3*m*, $\bar{4}$3*m*1', and $m\bar{3}m$. These Type-III altermagnets can exhibit some of other spin-split-relevant phenomena such as diagonal odd-order current-induced magnetization, off-diagonal odd-order current-induced magnetization, or even-order AHE…etc. One example of Type-III altermagnets is shown in Fig. 3j (MPG *mmm*), which has broken PT symmetry, but does not exhibit any of the phenomena listed above, at least, along any principal x/y/z axes. Diagonal odd-order current-induced magnetization along *z* requires broken $\{I\otimes T, I, M_x, M_y, M_x\otimes T, M_y\otimes T\}$. MPG *mmm* has unbroken $[R_x, R_y, R_z, M_x, M_y, M_z, I]$, so does not show diagonal odd-order current-induced magnetization along any principal x/y/z axes. However, the magnetic state of Fig. 3k (MPG 2221′ with crystallographic point group 422) has unbroken [$R_x, R_y, R_z, T, R_x \otimes T, R_y \otimes T, R_z \otimes T$] and broken {$I \otimes T, I, M_x, M_y, M_x \otimes T, M_y \otimes T$}, so does exhibit diagonal odd-order current-induced magnetization. Finally, we note that we have focused on ferromagnetic behaviors of altermagnets either in no external perturbations or in the presence of current or strain; however, spin splitting in altermagnets can be associated with other emerging physical phenomena with spin activities such as non-trivial superconductivity[15].

**CONCLUSION**

In the original proposal, a strong declaration like 'these materials should be viewed as members of a third magnetic class — altermagnets — alongside ferromagnets and antiferromagnets' was stated [9]. However, through symmetry considerations, we have demonstrated that rather than a fresh class, altermagnetism is a practically useful concept to introduce PT symmetry breaking and consequential ferromagnetic-like behaviors by combining crystallographic and spin alternations. The original altermagnets have fully-compensated 'collinear' spin angular momenta, but ferromagnet-like behaviors with non-zero orbital angular momenta through spin-orbital coupling with alternating two structural variations. We have extended the concept of altermagnetism to accommodate multiple director orientations (structural parameters, in general) and spin orientations, and this extended Altermagnetism includes 'non-collinear' spin states. In addition, we have classified three kinds of altermagnets; type-I - III. All type-I altermagnets, showing non-zero net magnetic moment and ferromagnetic-like behaviors, belong to ferromagnetic point group, and are a kind of weak ferromagnets. All type-II & III altermagnets, with the broken PT symmetry and zero net magnetic moment, can exhibit ferromagnet-like behavior in the presence of applied current or stress which are PT-symmetric. Type-II altermagnets exhibit high-odd order AHE, but Type-III altermagnets do not show any odd-order AHE. These various forms of altermagnetism offer an intriguing opportunity to leverage complementary advantages of both ferromagnetism and antiferromagnetism.

**ACKNOLEDGEMENT:** We thank Jairo Sinova, Allan H. Macdonald, Valery Kiryukhin, Seong Joon Lim for highly beneficial discussions on altermagnetism. The work at Rutgers University was supported by the DOE under Grant No. DOE: DE-FG02-07ER46382.

**ACKNOWLEDGEMENTS:** S.W.C. conceived and supervised the project. F.-T.H. conducted magnetic point group analysis. S.W.C. wrote the remaining part.

**COMPETING INTERESTS:** The authors declare no competing interests.

**DATA AVAILABILITY:** All study data is included in the article.

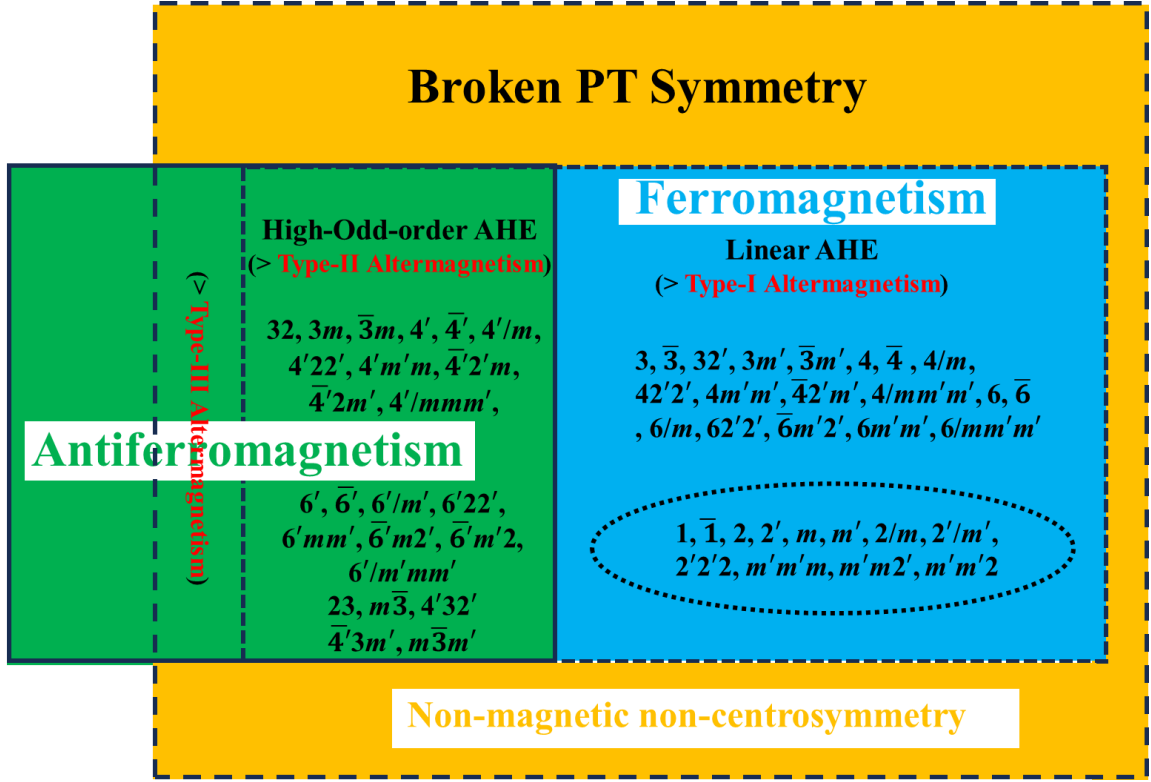

**FIGURE 1. Classification of various magnetic states.** All ferromagnets (blue box) belong to the ferromagnetic point group, and any magnetic states that do not belong to the ferromagnetic point group are antiferromagnetic (green box). All ferromagnets exhibit linear AHE, high-odd-order AHE can be observed in certain antiferromagnets, and all magnetic states, exhibiting odd-order AHE, have broken PT symmetry. Magnetic point groups relevant to type-I altermagnetism proposed for collinear spins are inside of dashed circle[18], and all type-I altermagnets belong to the ferromagnetic point group. All type-II altermagnets can exhibit high-odd-order AHE, and type-III altermagnets cannot show any odd-order AHE. Yellow area denotes non-magnetic and non-centrosymmetric states, which is not discussed in this perspective. Note that 'type-I altermagnetism' is a subset of 'ferromagnetism', 'type-II altermagnetism' is a subset of 'antiferromagnetism with high-odd-order AHE', and 'type-III altermagnetism' is a subset of 'antiferromagnetism without PT symmetry and odd-order AHE'; however, their exact subset relationships cannot be identified from symmetry analysis alone.

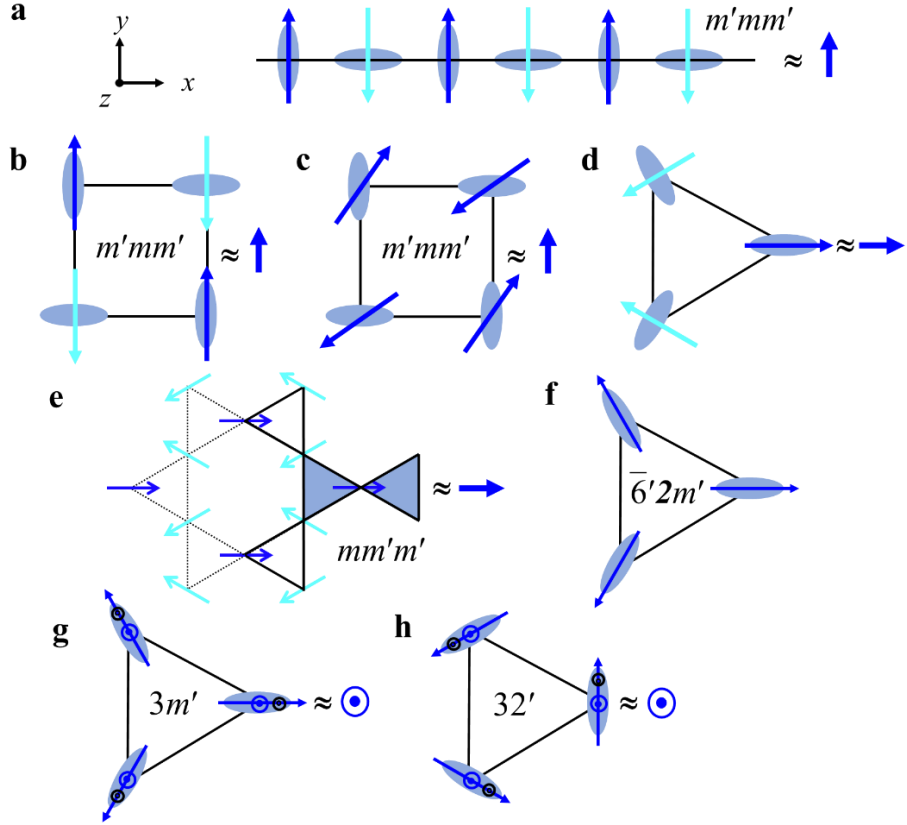

**FIGURE 2. Mostly Type-I altermagnets with non-zero net magnetic moment.** Various magnetic states with different orientations of local structures around one-kind magnetic ions (directors) and different orientations/magnitudes of ionic magnetic moments and their relevant net magnetic moments are shown. '≈' represents symmetry operational similarity. Blue arrows are magnetic moments, and the different blue tones of the blue arrows represent the different magnitudes of ionic magnetic moments even though they are from identical magnetic ions. **a** displays an alternating arrangement of 2 different director orientations and 2 different spin orientations in a chain. **b** displays an alternating arrangement of 2 different director orientations and 2 different spin orientations on a square lattice, and **c** shows another alternating arrangement of 2 directors and 2 different spin orientations on a square lattice. An alternating arrangement of 3 different director orientations and 3 different spin orientations on a triangle is displayed in **d**. **e** displays a spin configuration in kagome lattice with 120°-ordered spins and corresponds to the magnetic state of $Mn_3Ge$. **f** depicts an alternating arrangement of 3 different director orientations and 3 different spin orientations on a trimerized triangular lattice. This is a Type-II altermagnet with zero net magnetic moment. **g-h** show altermagnetic states with 3 different director

orientations and 3 different spin orientations in a trimerized triangular lattice – both directors and spins are tilted toward the out-of-plane direction. Black arrow heads denote out-of-plane director tilting.

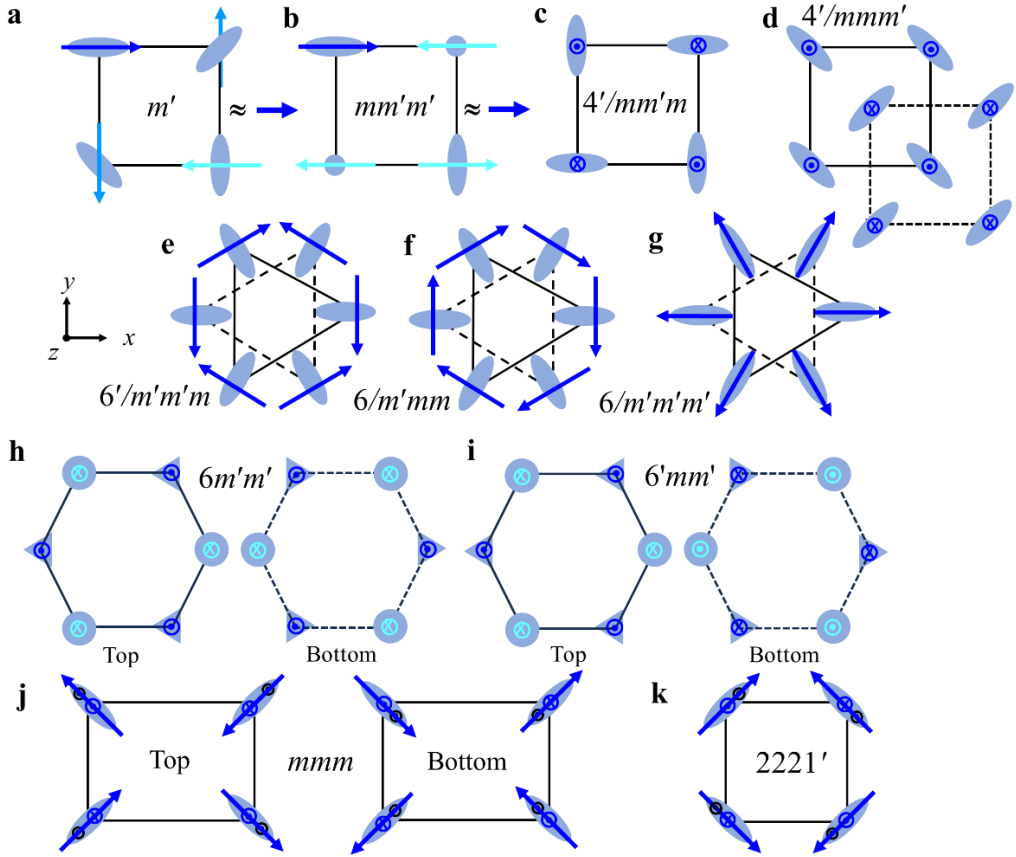

**FIGURE 3. Various altermagnets and non-altermagnets. a** depicts an altermagnetic state with 4 different director orientations and 4 different spin orientations in a square lattice, and **b** shows an altermagnetic state with 3 different director orientations and 2 different spin orientations in a square lattice. **c** displays an alternating arrangement of 2 different director orientations and 2 different spin orientations in a square lattice and is not associated with any net magnetic moment. **d** denotes AB-stacked square lattices with alternating director and spin orientations in different layers. **e-g** display alternating arrangements of 3 different director orientations and 3 different spin orientations on an AB-stacked trimerized triangle lattice (or hexamerized honeycomb lattice). **h-i** show alternating arrangements of tetrahedral (triangular symbols) and octahedral (circular symbols) coordinates and up-down magnetic moments on stacked honeycomb lattice. One apex of each tetrahedral coordinate is always pointing down. **h** is associated with non-zero net magnetic moment along *z*, but **i** has no net magnetic moment. **j** displays alternating arrangements of 4 different director orientations and 4 different spin orientations on an AA-stacked rectangular lattice. Black arrow heads in **j** and **k** denote out-of-plane director tilting. **k** displays alternating arrangements of 4 different director orientations and 4 different spin orientations and is not associated with any net magnetic moment. All of the magnetic states, except **f-g**, are altermagnets with the broken PT symmetry. **a**, **b**, & **h** are type-I altermagnets with non-zero net magnetic moment, **c-e** and **i** are type-II altermagnets with zero net magnetic moment. **j** is type-III altermagnets with zero net magnetic moment and also zero odd-order AHE. **k** is a type-III altermagnets with diagonal odd-order current-induced magnetization. f-g are non-altermagnets.